\documentclass[aps,prl,twocolumn,superscriptaddress,longbibliography]{revtex4}
\usepackage{graphicx}
\usepackage{dcolumn}
\usepackage{bm}
\usepackage{physics}
\usepackage{amsmath}
\usepackage{amssymb}
\usepackage{array}
\usepackage{soul}
\usepackage{color}
\usepackage{xcolor} 
\newcolumntype{P}[1]{>{\centering\arraybackslash}p{#1}}
\usepackage[colorlinks=true, breaklinks=true, linkcolor=blue, citecolor=blue, urlcolor=blue]{hyperref}

\newcommand{\bonnpi}{Physikalisches Institut, University of Bonn, Nussallee 12, 53115 Bonn, Germany}
\newcommand{\geneva}{Department of Quantum Matter Physics, University of Geneva, Quai Ernest-Ansermet 24, 1211 Geneva, Switzerland}
\newcommand{\saarbrucken}{Theoretische Physik, Universit\"at des Saarlandes, Campus E26, D-66123 Saarbr\"ucken, Germany}

\begin{document}

\title{From Light-Cone to Supersonic Propagation of Correlations\\ by Competing Short- and Long-Range Couplings}
\date{\today}

\begin{abstract}
We investigate the dynamical spreading of correlations in many-body quantum systems with competing short- and global-range couplings. 
We monitor the non-equilibrium dynamics of the correlations following a quench, showing that for strong short-range couplings the propagation of correlations is dominated at short and intermediate distances by a causal, light-cone, dynamics, resembling the purely short-range quantum systems.
However, the interplay of short- and global-range couplings leads to a crossover between space-time regions in which the light-cone persists to regions where a supersonic, distance-independent, spreading of the correlations occurs. 
We identify the important ingredients needed for capturing the supersonic spreading and demonstrate our findings in systems of interacting bosonic atoms, in which the global range coupling is realized by a coupling to a cavity light field, or atomic long-range interactions, respectively. 
We show that our results hold in both one and two dimensions and in the presence of dissipation.
Furthermore, we characterize the short time power-law scaling of the distance-independent growth of the density-density correlations.  
\end{abstract}

\author{Catalin-Mihai Halati}
\affiliation{\geneva}
\author{Ameneh Sheikhan}
\affiliation{\bonnpi}
\author{Giovanna Morigi}
\affiliation{\saarbrucken}
\author{Corinna Kollath}
\affiliation{\bonnpi}
\author{Simon B.~J\"{a}ger}
\affiliation{\bonnpi}
\maketitle

In recent decades, a fundamental question attracting enormous interest is what are the speed limits on how fast physical effects can propagate in non-relativistic quantum systems  \cite{LiebRobinson1972, ChenYin2023, DefenuPappalardi2024, DefenuTrombettoni2023}.
Beside its fundamental character, this problem is also relevant for quantum technologies, as it bounds the timescale of quantum operations and algorithms \cite{DeffnerCampbell2017}.
The knowledge into how correlations propagate enables one to select the most suitable platform to engineer the dynamical features leading to the desired speed.

In quantum systems with short-range interactions Lieb and Robinson have shown that correlations can only travel within a light-cone defined by a maximal velocity \cite{LiebRobinson1972}. 
The Lieb-Robinson bound is a key concept across different topics in quantum many-body physics, quantum thermodynamics, and quantum information \cite{ChenYin2023}, e.g.~the exponential clustering for correlation functions \cite{NachtergaeleSims2006}, the scrambling of information \cite{RobertsSwingle2016}, and even the area law of entanglement \cite{Hastings2007, VanAcoleyenVerstraete2013}.
Although the original derivation of this bound is not strictly applicable to bosonic systems, an effective light-cone can be derived at low densities \cite{TakasuTakahashi2020,KuwaharaSaito2021, YinLucas2022} and
in the Bose-Hubbard model \cite{KollathAltman2007, LaeuchliKollath2008, CheneauKuhr2012, BarmettlerKollath2012, CarleoFabrizio2014,FaupinSigal2022,FaupinSigal2022b}, while the spreading of correlations breaking the light-cone bound is typically only present in tailored models \cite{EisertGross2009, KuwaharaSaito2024, WangHazzard2020}. 
The typical light-cone spreading of correlations for short-range couplings has been observed experimentally with ultra-cold atoms in optical lattices \cite{CheneauKuhr2012, BarmettlerKollath2012}.

The extension of these concepts to long-range interacting systems, where the interactions can directly connect distant sites, is non-trivial and partially counterintuitive \cite{DefenuPappalardi2024, DefenuTrombettoni2023}.
An important class of long-range interactions are algebraically decaying with the distance, i.e.~$x^{-\alpha}$, naturally occurring in trapped ions, Rydberg, and dipolar systems \cite{DefenuTrombettoni2023}. 
The exponent can also be tailored by engineering interactions with lasers and resonators \cite{PeriwalSchleier-Smith2021, SchneiderSchaetz2012}. 
While, in general, the concept of Lieb-Robinson bounds is not applicable to arbitrary long-range interacting systems~\cite{Hastings2012, EisertKastner2013, Foss-FeigGorshkov2015, SchneiderSanchez-Palencia2021, VuSaito2024}, an effective bound exists for sufficiently large values of $\alpha$ \cite{EisertKastner2013,  HaukeTagliacozzo2013, CevolaniSanchez-Palencia2015, ChenLucas2019, TranLucas2020, KuwaharaSaito2020, Foss-FeigGorshkov2015, ElseYao2020, ChenLucas2021, TranLucas2021, KuwaharaSaito2021b,FaupinSigal2022,FaupinSigal2022b, GongCirac2023, VuSaito2024}, or special setups \cite{JunemannGarcia-Ripoll2013}.
The spreading of correlation has been experimentally measured in trapped ion chains with tailored interactions \cite{RichermeMonroe2014, JurcevicRoos2014}, reporting the expected deviations from the light-cone propagation \cite{Foss-FeigGorshkov2015, ChenLucas2021, TranLucas2021}.

For the special case of strong long-range interactions, where $\alpha$ is smaller than the spatial dimension, theoretical bounds allow for correlations propagation that can be almost instantaneous \cite{TranLucas2020}. On the other hand, for certain initial states, local perturbations can remain frozen, exhibiting effects analogous to localization \cite{SantosCelardo2016, CelardoBorgonovi2016} originating from large energy gaps in the spectra of strong long-range interacting systems. 
Remarkably, little is known on how correlations spread in systems with competing short- and long-range interactions. While previous work shed light on the intricate thermalization behavior in the presence of long- and short-range interactions \cite{BlassRieger2018, HalimehKastner2017, ZunkovicSilva2018, LeroseSilva2019, DefenuPappalardi2024}, theoretical analyses of correlations spreading in this situation relevant for state-of-the art experiments are rare.

In this work, we investigate the dynamical spreading of correlations after a quench in a system with competing global- and short-range couplings.
We consider a paradigmatic model of experimental relevance, namely, the Bose-Hubbard model of cavity quantum electrodynamics \cite{RitschEsslinger2013, MivehvarRitsch2021}, where the short-range kinetic and on-site processes compete with global density-density interactions mediated by photon scattering. 
We perform numerically exact simulations based on matrix product states and supplement them with approximate numerical and analytical techniques.
We characterize the interplay of light-cone dynamics and of the non-causal propagation of the globally interacting dynamics, reporting a crossover between the light-cone behavior and the supersonic spreading of correlations, where correlations can spread across the system almost simultaneously, independent of the distance. 
We also show that the supersonic spreading can be triggered by light-cone dynamics.
While the exact simulations are crucial in obtaining the details of the non-equilibrium dynamics, the approximate approaches allow us to identify the key ingredients necessary for the supersonic propagation and to pinpoint fluctuations of the global coupling which act as the carriers of the long-range correlations.

\emph{Setup:}
We consider bosonic atoms in a lattice, which interact with a quantum field, representing a photonic mode mediating global interactions between the atoms.
The evolution of the density matrix of the coupled system, $\hat{\rho}$, is given by the Lindblad master equation \cite{CarmichaelBook, BreuerPetruccione2002, MaschlerRitsch2008, RitschEsslinger2013, MivehvarRitsch2021}
\begin{align}
\label{eq:Lindblad}
& \pdv{t} \hat{\rho} = -\frac{i}{\hbar} \left[ \hat{H}, \hat{\rho} \right] + \frac{\Gamma}{2}\left(2\hat{a}\hat{\rho} \hat{a}^\dagger-\hat{a}^\dagger \hat{a} \hat{\rho}-\hat{\rho} \hat{a}^\dagger \hat{a}\right), 
\end{align}
where we include the losses of the cavity with rate $\Gamma$. The Hamiltonian reads 
\begin{align} 
\label{eq:Hamiltonian}
&\hat{H}=\hat{H}_{\text{cav}}+\hat{H}_{\text{BH}},~\hat{H}_{\text{cav}}= \hbar\delta \hat{a}^\dagger \hat{a} -\hbar\Omega ( \hat{a} + \hat{a}^\dagger)\hat{\Delta},\\
&\hat{H}_{\text{BH}}=\frac{U}{2} \sum_{j} \hat{n}_{j}(\hat{n}_{j}-1)-J \sum_{\langle j,j'\rangle} (\hat{b}_{j}^\dagger \hat{b}_{j'} + \text{H.c.}). \nonumber
\end{align}
The local processes consist in the repulsive on-site interactions of strength $U$ and atomic tunneling between neighboring sites $\langle j,j'\rangle$ with the amplitude $J$. The model provides a very good description of the dynamics realized in experiments \cite{KlinderHemmerich2015b, LandigEsslinger2016, HrubyEsslinger2018}, where the atomic transition is far detuned and spontaneous decay can be neglected.
The effective light-matter coupling strength $\Omega$ is controllable by a pump laser and $\delta$ is the cavity-pump detuning. 
The period of the lattice is twice the period of the cavity mode coupling the cavity to the density imbalance $\hat{\Delta}=\sum_{j\in A} \hat{n}_j-\sum_{j\in B} \hat{n}_j$ of a bipartite lattice with sublattices $A$ and $B$ \cite{HabibianMorigi2013b, NiederleRieger2016, LandigEsslinger2016}.
Theoretical efforts characterized mostly the nature of steady states in such models \cite{NiedenzuRitsch2010, SilverSimons2010, VidalMorigi2010,LiHofstetter2013, BakhtiariThorwart2015,FlottatBatrouni2017, LinLode2019, HimbertMorigi2019, HalatiKollath2020, BezvershenkoRosch2021, SharmaMorigi2022, ChandaMorigi2022}, and certain aspects of their dynamics \cite{ChiacchioNunnenkamp2018, HalatiKollath2022, HalatiKollath2025}.

\begin{figure}[!hbtp]
\centering
\includegraphics[width=0.48\textwidth]{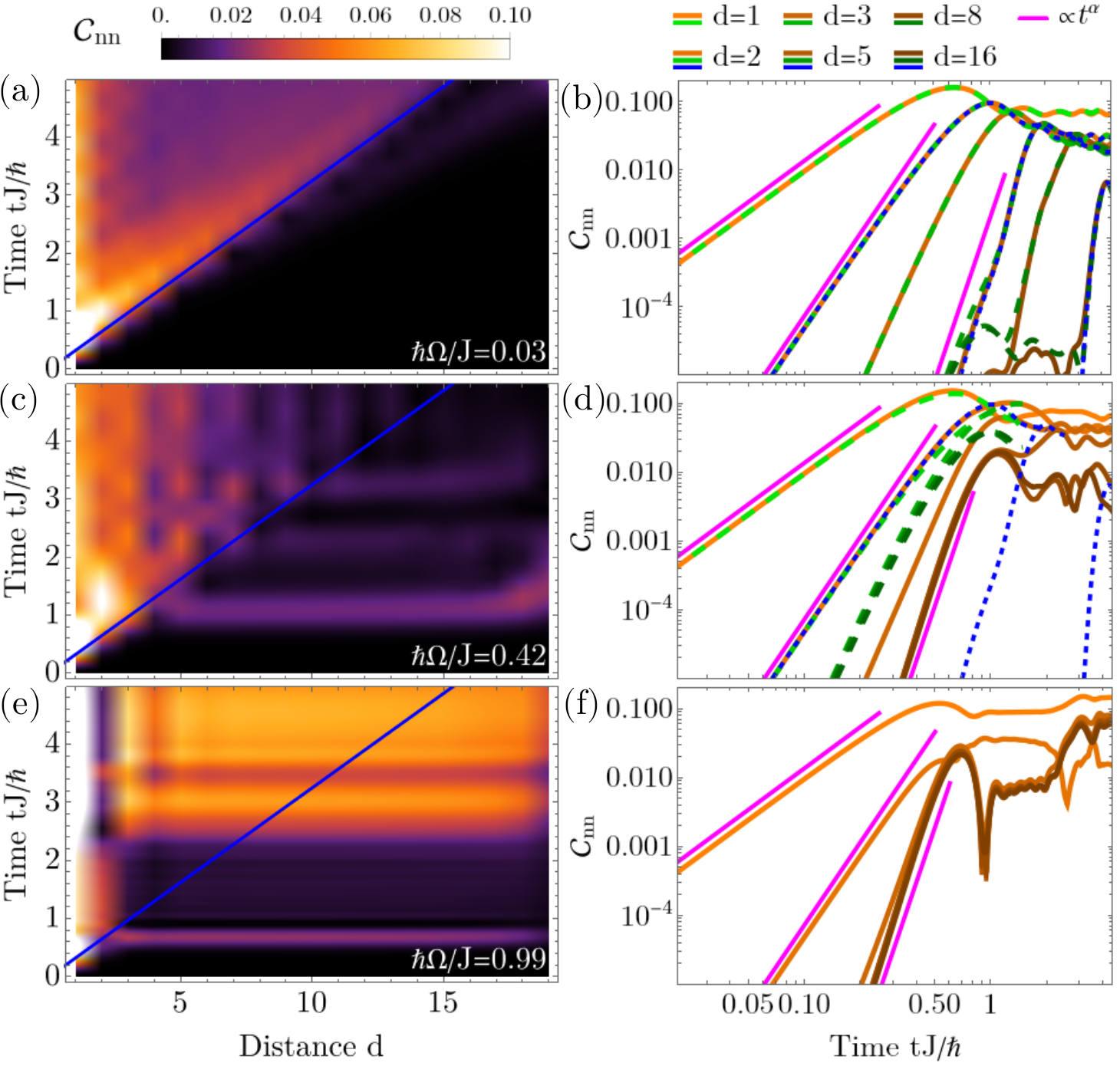}
\caption{(a), (c), (e) The propagation of $\mathcal{C}_{nn}(d,t)$, computed with $\hat{H}$, Eq.~\eqref{eq:Hamiltonian}, for the atoms-cavity coupling strength $\hbar\Omega/J\in\{0.03,0.42,0.99\}$. The blue line is a guide to the eye, approximating the front of the light-cone propagation for $\Omega=0$.
(b), (d), (f) $\mathcal{C}_{nn}(d,t)$ for several distances and $\hbar\Omega/J\in\{0.03,0.42,0.99\}$.
The curves depicted with shades of orange to brown correspond to $\hat{H}$, Eq.~\eqref{eq:Hamiltonian}, the dashed curves with shades of green to $\hat{H}_{\text{atom-only}}$, Eq.~\eqref{eq:Hamiltonian_atom_only}, and the dashed blue curves to the $\Omega=0$ case.
The magenta lines show the algebraically scaling $\propto t^\alpha$, with $\alpha\in\{2,4,8\}$.
The parameters used are $N=10$ particles, $L=20$ sites, $U/J=2$, $\hbar\delta/J=2$, $\hbar\Gamma/J=0$.
}
\label{fig:lightcone}
\end{figure}

\emph{Quench scenario:}
We simulate the propagation of correlations in a quench scenario in which initially the atoms are in an uncorrelated product state with one atom every two sites, $\ket{1010\dots}$, and the cavity is empty. The initial state breaks the $\mathbb{Z}_2$ symmetry of $\hat{H}$, $(\hat{a},\hat{\Delta})\!\to\!(-\hat{a},-\hat{\Delta})$.
We analyze the dynamics of connected density-density correlations
\begin{align} 
\label{eq:correlations}
\mathcal{C}_{nn}(d,t)=\frac{1}{\mathcal{N}}\sum_{j,j',|j-j'|=d}\left|\left\langle \hat{n}_j \hat{n}_{j'}\right\rangle-\left\langle \hat{n}_j\right\rangle \left\langle \hat{n}_{j'}\right\rangle\right|(t),
\end{align}
where we average over all correlations at a certain distance $d\equiv|j-j'|$ with $\mathcal{N}$ being the number of sites at the given distance $d$. 
In $\mathcal{C}_{nn}(d,t)$ we subtract the disconnected part of the density-density correlations, $\left\langle \hat{n}_j\right\rangle \left\langle \hat{n}_{j'}\right\rangle$, 
which describes the reorganization of the atomic density in the cavity-induced potential.

As in the initial state the different sites are uncorrelated, this scenario allows us to investigate cleanly the space-time propagation of correlations  through the system. 
We remark that one particularity of the considered correlations is that in the limit of vanishing tunneling, $J\!=\!0$, $\hat{H}$ commutes with density operators and the correlations $\mathcal{C}_{nn}$ would be constant in time. 
We investigate the propagation of correlations for $J\!\neq\!0$, in the presence of competing short- and global-range terms for the 1D version of $\hat{H}$, Eq.~\eqref{eq:Hamiltonian}, and of the Liouvillian, Eq.~\eqref{eq:Lindblad}, using a recently developed method based on time-dependent matrix product states (tMPS) employing swap gates for the cavity coupling and quantum trajectories for the dissipative dynamics \cite{HalatiKollath2020b, supp}

\emph{Light-cone to supersonic evolution:}
We investigate the crossover between the light-cone evolution and distance-independent propagation of the correlations in Fig.~\ref{fig:lightcone}, for the 1D chain of atoms coupled to the cavity, assuming no photon losses ($\Gamma\!=\!0$). 
For low coupling to the cavity field [Fig.~\ref{fig:lightcone}(a) for $\hbar\Omega/J=0.03$] and the considered distances a light-cone propagation is found following the blue line, which marks approximately the speed limit for correlations spreading we found in the absence of the global-range coupling.
For the same value of the coupling, we present in Fig.~\ref{fig:lightcone}(b)  $\mathcal{C}_{nn}(d,t)$ for certain distances as a function of time. 
The correlations exhibit an algebraic increase with time until reaching a  maximum following the light-cone. 
By comparing with the $\Omega=0$ case we observe for correlations at longer distances, $d \gtrsim 10$, finite values outside of the light-cone, signaling the supersonic spreading. The light-cone spreading at small times and distances can be understood by expanding the time-evolution operator, i.e.~$e^{-i\hat{H}t/\hbar} \!\approx\!  e^{-i\hat{H}_{\text{BH}}t} +O(t^2\Omega)$. Thus, the time-evolution is dominated by the Bose-Hubbard evolution beside corrections on the order of $O(t^2\Omega)$, or higher depending on the initial state and the form of the coupling \cite{supp}.

The features of the supersonic propagation become apparent for larger values of the coupling strength, $\hbar\Omega/J=0.42$ in Fig.~\ref{fig:lightcone}(c)-(d). 
Here, we observe a coexistence of the light-cone induced by short-range processes and a distance-independent increase of correlations, for $d>4$, due to the presence of the cavity-induced global-range interactions.
In Fig.~\ref{fig:lightcone}(d) we observe that $\mathcal{C}_{nn}(d,t)$ follows the light-cone at very short distances and times ($tJ/\hbar\lesssim 1$, $d\leq3$), while for $d\geq 5$ all correlations increase simultaneously.
This is due to density correlations commuting with the entire Hamiltonian except the tunneling terms, implying that an initial change in the correlations can only be induced by tunneling. 
Thus, the cavity-mediated long-range interactions can start spreading the correlations only triggered by the short-range hopping term. 
By increasing the coupling strength to the cavity even further, $\hbar\Omega /J\!=\!0.99$ in Fig.~\ref{fig:lightcone}(e)-(f), we observe that the instantaneous spreading of the correlations occurs at a shorter times and only for $d\leq2$ the correlations grow beforehand.
Furthermore, for such strong long-range interactions we do not observe the light-cone dynamics, the role of the short-range processes being restricted to generating locally the density-density correlations, which propagate via the cavity coupling.
After the initial rise, an intricate oscillatory behavior is observed in space and time, stemming from the interplay of the Bose-Hubbard terms \cite{BarmettlerKollath2012}, and the odd-even global ordering induced by $\hat{\Delta}$.
This behavior could be related to metastability typical of global-range interacting systems \cite{SantosCelardo2016, ChavezCelardo2021, DefenuPappalardi2024} and its study is left for future work.

\begin{figure}[!hbtp]
\centering
\includegraphics[width=0.48\textwidth]{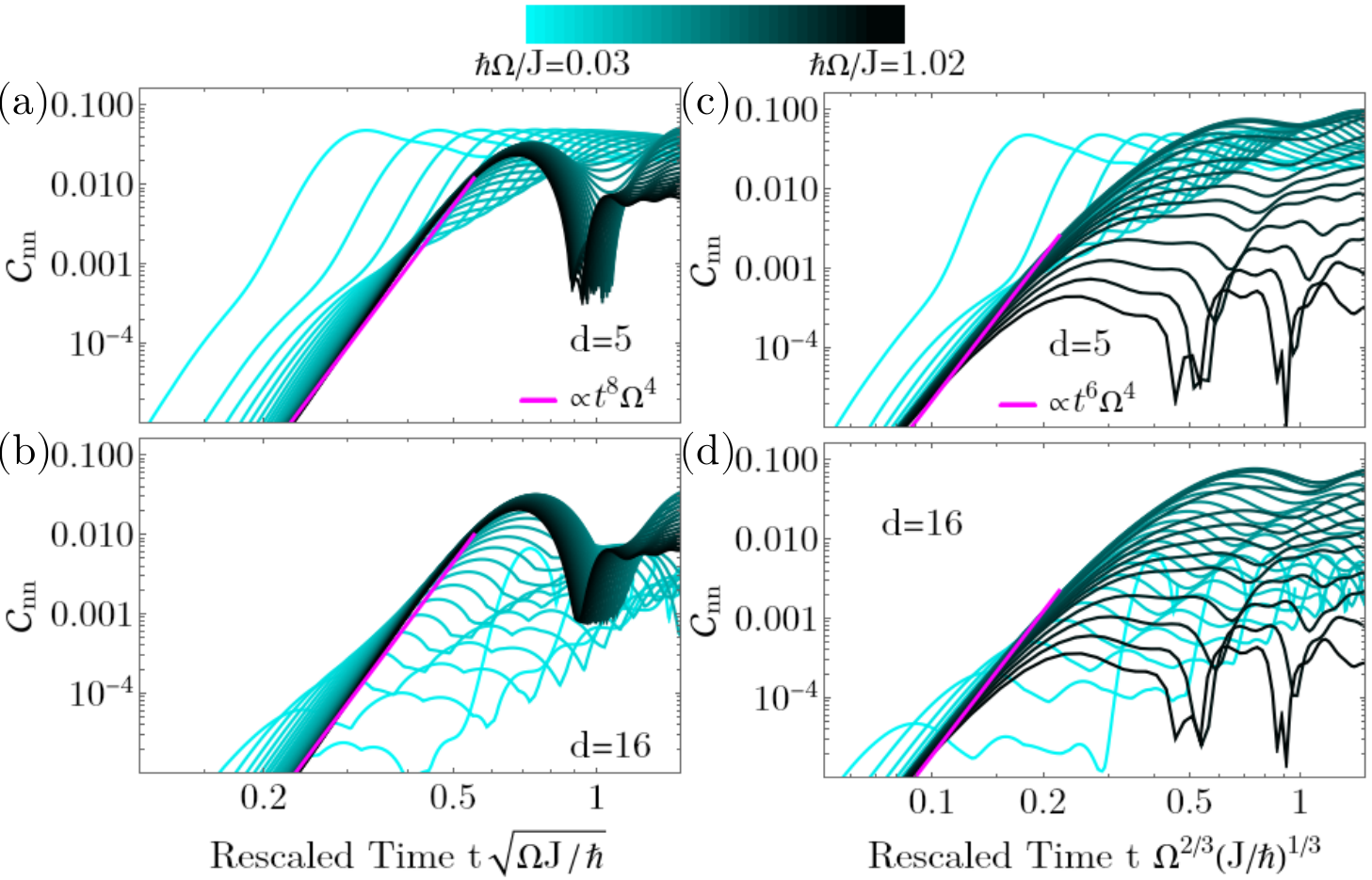}
\caption{(a)-(b) The dependence of the correlations $\mathcal{C}_{nn}$ as a function of rescaled time $t\sqrt{\Omega J/\hbar}$ for (a) $d=5$ and (b) $d=16$, obtained for the atom-cavity Hamiltonian $\hat{H}$, Eq.~\eqref{eq:Hamiltonian}.
(c)-(d) The dependence of the correlations $\mathcal{C}_{nn}$ as a function of rescaled time $t\Omega^{2/3} (J/\hbar)^{1/3}$ for (c) $d=5$ and (d) $d=16$, obtained for the atom-only Hamiltonian $\hat{H}_{\text{atom-only}}$, Eq.~\eqref{eq:Hamiltonian_atom_only}.
The magenta lines represent the algebraic scaling (a)-(b) $\propto t^8 \Omega^4$ and (c)-(d) $\propto t^6 \Omega^4$.
The values of the coupling are $0.03\leq\hbar\Omega/J\leq1.02$, and the same parameters as in Fig.~\ref{fig:lightcone}.
}
\label{fig:scaling}
\end{figure}

\emph{Scaling of the supersonic spreading:}
The short-time behavior of the distance-independent spreading can be understood with the minimal model of a simplified tunneling term and the global coupling to the cavity \cite{supp}. 
The correlations build up following the scaling
\begin{align} 
\label{eq:scaling}
\mathcal{C}_{nn}(d,t)\!\propto\! \Omega^4 J^4 t^8 \left[\langle (\hat{a}+\hat{a}^\dagger)^2\rangle-  \langle \hat{a}+\hat{a}^\dagger\rangle^2\right]
\end{align}
for $\hbar\Omega\!\gg\!J$ and an initial uncorrelated Fock state. 
Here the amplitude is proportional to the cavity field fluctuations, implying that the fluctuations are crucial to induce the supersonic spreading.
In Fig.~\ref{fig:scaling}(a),(b) we show that this scaling is dominant in the large coupling regime of the atoms-cavity model $\hat{H}$ by collapsing 
the numerical results for a wide range of different values of $\Omega$ onto a single curve using $(t\sqrt{\Omega J})^8$.
This scaling changes when other energy scales, such as the full kinetic term, the detuning $\delta$, or preexisting long-range density correlations in the initial state, become relevant.
The curves at lower $\Omega$ values deviate from the scaling gradually, showing a gradual change between the light-cone and supersonic propagation of the correlations. 
Thus, the interplay between the kinetic process and the global coupling to the cavity field is sufficient to cause the crossover, however, the global coupling fluctuations are required in this minimal model to obtain the supersonic spreading.

\emph{Propagation mediated by atom-only global interactions:}
In the full model, Eq.~\eqref{eq:Hamiltonian}, the cavity field acts as an explicit carrier of the long-range correlations. The question arises if a global purely atomic interaction leads to similar behavior.
Thus, we contrast the results of the coupled atoms-cavity system with the atom-only model, 
\begin{align} 
\label{eq:Hamiltonian_atom_only}
&\hat{H}_\text{atom-only}=\hat{H}_{\text{glo}}+\hat{H}_{\text{BH}},~\hat{H}_{\text{glo}}=  -\frac{\hbar \Omega^2 \delta}{\delta^2+\Gamma^2/4} \hat{\Delta}^2.
\end{align}
This represents an atom-only description of the complex hybrid system obtained by eliminating the cavity field in the limit $ J,U,\hbar\Omega^2/\delta\ll\hbar\delta$ \cite{RitschEsslinger2013, MivehvarRitsch2021}, and can include global range dissipative processes \cite{VidalMorigi2010,HabibianMorigi2013b, MivehvarRitsch2021, ChiacchioNunnenkamp2018, JagerBetzholz2022}.
$\hat{H}_\text{atom-only}$ includes directly the effective global-range nature induced by the coupling to the cavity in $\hat{H}_{\text{glo}}$. 
The simulations of the atom-only model, Eq.~\eqref{eq:Hamiltonian_atom_only}, make use of the matrix product states implementation of the time-dependent variational principal (TDVP) \cite{HaegemanVerstraete2011, HaegemanVerstraete2016, supp}.
We show that even for a different global-range coupling the same crossover in the correlation spreading occurs as for the coupling to the cavity field.

In Fig.~\ref{fig:lightcone}(b),(d), showing the dynamics of $\mathcal{C}_{nn}(d,t)$ for several distances, we plot both the results for the atoms-cavity Hamiltonian, Eq.~\eqref{eq:Hamiltonian}, (continuous lines), and the results for the corresponding atom-only model, Eq.~\eqref{eq:Hamiltonian_atom_only}, (dashed lines).
At small values of the long-range interactions, Fig.~\ref{fig:lightcone}(b), we obtain a very good quantitative agreement in the dynamics of $\mathcal{C}_{nn}(d,t)$.
In Fig.~\ref{fig:lightcone}(d) we observe that the initial rise of the correlations $d=1$ and $d=2$ agrees in the two models, confirming that they are mostly influenced by the short-range terms. While in both cases we have a simultaneous increase of $\mathcal{C}_{nn}(d>3,t)$, the rise of $\mathcal{C}_{nn}(d>3,t)$ for the atom-only model occurs at slightly earlier times than for the atoms-cavity model. 
In particular, the scaling we find is following more closely $ \mathcal{C}_{nn}(d,t)\!\propto\!\Omega^4 t^6$, as shown in Fig.~\ref{fig:scaling}(c),(d). 
We note that by decreasing the photonic time-scale, i.e.~increasing $\delta$, while keeping $\Omega^2/\delta$ constant, the difference between the long distance correlations of the two models decreases (not shown) and eventually they agree in the limit in which $\hbar\delta$ is the largest energy scale. 
In this regime also the atoms-cavity model shows the scaling $ \mathcal{C}_{nn}(d,t)\!\propto\!\Omega^4 t^6$, implying that the main effect of the fast photons is to introduce the effective global interaction. 
The qualitative agreement in the behavior of the correlations in the two models shows that the main features of the competition of the light-cone and the distance-independent propagation due to the global range interactions are more generic than the case of hybrid atoms-cavity systems.

\begin{figure}[!hbtp]
\centering
\includegraphics[width=0.48\textwidth]{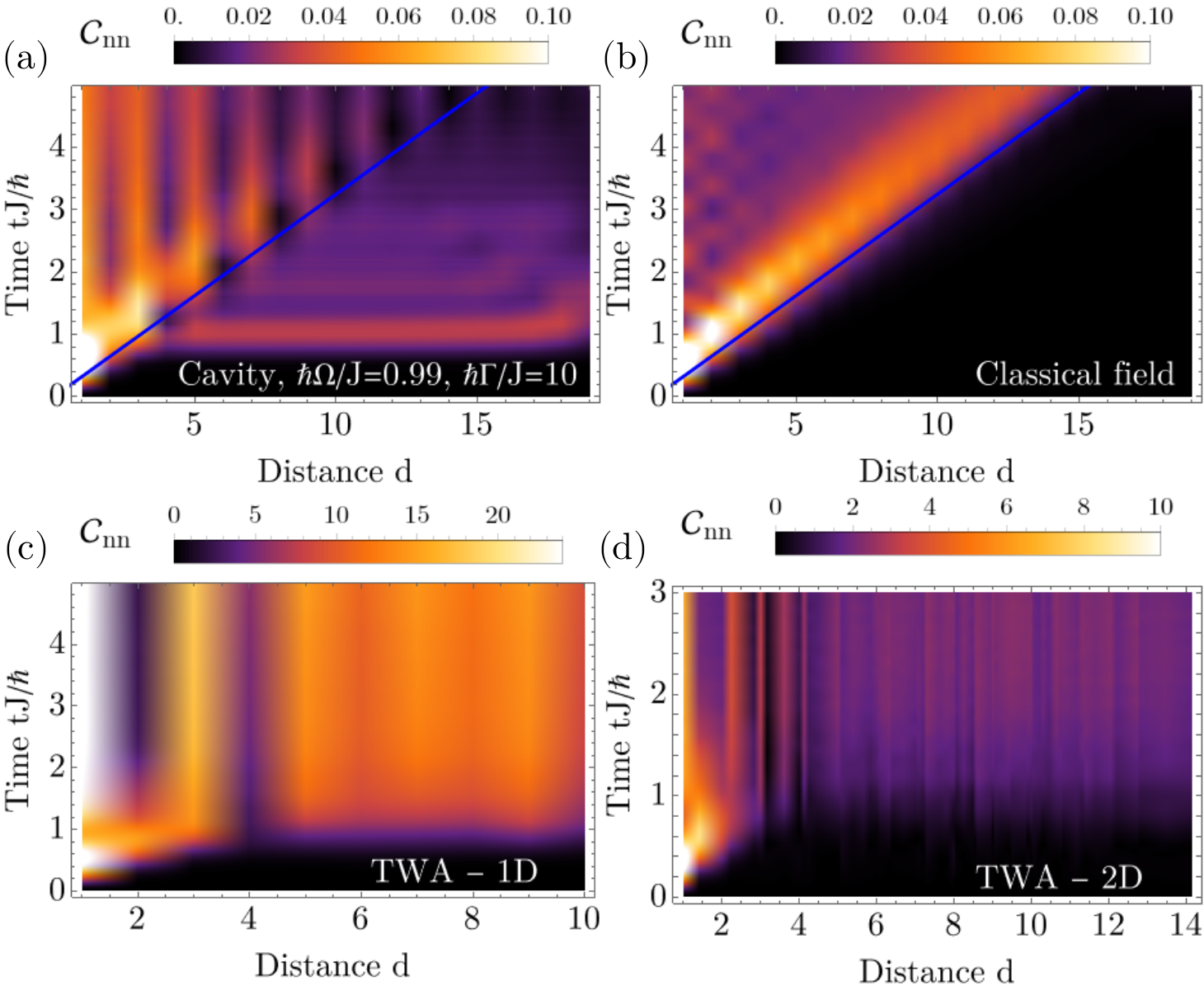}
\caption{(a) The propagation of $\mathcal{C}_{nn}(d,t)$, in the presence of dissipation, Eqs.~(\ref{eq:Lindblad})-(\ref{eq:Hamiltonian}), for $\hbar\Omega/J=0.99$, $\hbar\Gamma/J=10$, $N=10$ particles, $L=20$ sites, $U/J=2$, $\hbar\delta/J=2$. The blue line is a guide to the eye, approximating the front of the light-cone propagation for $\Omega=0$ and $\Gamma=0$.
(b) The propagation of correlations when cavity field is replaced with a classical field \cite{supp}, for the same parameters as in (a). 
Correlations $\mathcal{C}_{nn}(d,t)$ calculated from TWA in (c) 1D for a lattice with L=21 and in (d) 2D for lattice with $L\times L=21\times21$. For the TWA simulations we have used $U/J=0.1$, $\hbar\delta/J=2$, $\hbar\Gamma/J=1$, $\sqrt{N}\hbar\Omega/J=4$ in (c), and $\sqrt{N}\hbar\Omega/J=8$ in (d). We initialize the atoms with alternating densities of $n=10$ and $n=11$ bosons corresponding to the total atom numbers (c)~$N=220$ and (d)~$N=4630$.
}
\label{fig:dissipation}
\end{figure}

\emph{Influence of a dissipative cavity field:}
A seldom explored question is how the presence of dissipation alters the spreading of correlations.
Seminal works showed that for quasi-local Lindblad operators Lieb-Robinson-like bounds  exist \cite{Poulin2010, SchuchEisert2011, BarthelKliesch2012, CampoHuelga2013, FunoSaito2019}, observed these bounds numerically \cite{BernierKollath2018} and characterized entanglement measures \cite{AlbaCarollo2021, WellnitzSchachenmayer2022}.
The dissipation range can further alter the spatio-temporal behavior of correlations \cite{SeetharamMarino2022}.
Thus, we aim to understand the effects of dissipation in form of cavity losses, Eq.~\eqref{eq:Lindblad}, which is a highly non-local dissipation for the atoms due to the global coupling.

For the atoms-cavity system, Eqs.~\eqref{eq:Lindblad}-\eqref{eq:Hamiltonian}, photonic dissipation alters the position of the crossover between the local and the global spreading of correlations. 
By increasing the strength of dissipation we go from the regime in which the supersonic propagation dominates to a regime with a mostly light-cone evolution [see Fig.~\ref{fig:dissipation}(a) compared to Fig.~\ref{fig:lightcone}(e)-(f)]. 
Thus, whereas the dissipation globally couples to the atoms, 
the photon losses rescale the effective global atom-atom interactions to a lower value [see the coefficient in $\hat{H}_\text{glo}$ Eq.~\eqref{eq:Hamiltonian_atom_only}], making the light-cone propagation more prominent. However, for the same strength of $\hat{H}_\text{glo}$ stronger dissipation helps the supersonic propagation.

\emph{Key ingredients for the supersonic propagation:}
Our results show the presence of the supersonic propagation for global coupling to the cavity and atom-only global interactions. 
Thus, we aim to identify the underlying carrier of the correlations, using different approximations to highlight the required ingredients. 
In Fig.~\ref{fig:dissipation}(b) we show results obtained employing a mean-field description of the cavity field, for which the cavity mode is modeled by a time-dependent classical field coupled to the mean value $\langle\hat{\Delta}\rangle$ \cite{supp}.
In this mean-field approach we can only recover the light-cone evolution while the distance-independent spreading of correlations is absent [see Fig.~\ref{fig:dissipation}(b)]. This is due to the inability of the cavity in the mean-field description to transport and create fluctuations. Note that also in the minimal model, Eq.~\eqref{eq:scaling}, the amplitude of the supersonic spreading would vanish in the mean-field approximation. 
Thus, we identify fluctuations, which can have different origins, as the carriers of the supersonic correlations.
One source of fluctuations in the cavity field is the stochastic noise arising from photon losses. By including such a term in the mean-field description, which translates to a correlated noise in the atomic evolution, we recover a distance-independent propagation, albeit with a reduced contribution than in Fig.~\ref{fig:dissipation}(a) \cite{supp}.

To analyze the role of the fluctuations arising in the atoms-cavity coupling we employ the truncated Wigner approximation (TWA).
In the TWA the cavity field and each atomic site is described as a complex stochastic classical field \cite{GardinerFudge2002, supp} and it has been previously shown to describe light-cone spreading of correlations in Bose-Hubbard models~\cite{NagaoDanshita2019}.
For the 1D case at high filling, we show in Fig.~\ref{fig:dissipation}(c) the TWA results, for an initial state with a staggered atomic occupation alternating between $n\!=\!10$ and $n\!=\!11$ \cite{supp}.
We highlight that the qualitative behavior is the same as discussed in Fig.~\ref{fig:lightcone}(e) for the tMPS results: at short times light-cone spreading of the correlations occurs, followed by a distance-independent rise of correlations at large distances. 
Thus, the fluctuations of the global coupling through the cavity field determine the supersonic spreading.

\emph{Spreading of correlations in two-dimensions:}
Establishing the TWA as a valid approach to simulate the quantum dynamics of correlations in this model gives us the opportunity to study this effect in higher dimensions.
In Fig.~\ref{fig:dissipation}(d), we show the results for a 2D atomic system coupled to the cavity field, where the atomic initial state is a checkerboard with alternating $n\!=\!10$ and $n\!=\!11$ occupations.
Compared to the 1D simulations the magnitude of $\mathcal{C}_{nn}$ is reduced, which we attribute to radial spread of the correlations in 2D. 
Nevertheless, the key features that emerge from the competing short- and long-range processes are well visible in 2D, the light-cone propagation of correlations on short times and the supersonic spreading on longer times.

\emph{Conclusions:}
We investigated the dynamics of density-density correlations emerging from the interplay of short- and global-range couplings, finding two key features, a light-cone propagation at short distances and times followed by a supersonic, distance-independent, spreading. 
The crossover time between these dynamical features and their relative strength depend crucially on the ratio between the global coupling and the short-range tunneling. 
The general character of our results is highlighted by the robustness to the different quench situations, the dimensionality of the system, the presence of dissipation, or the form of the global coupling.
We further checked the stability of the features under local atomic dephasing~\cite{supp}.
We identify that the emerging dynamics relies on the presence of fluctuations in the long-range couplings, stemming either from the cavity field or the interaction itself. 
The role of fluctuations is essential for propagating the correlations outside of the light-cone, which further underline the importance of capturing the fluctuations in the atoms-cavity coupling in obtaining the correct quantum dynamics \cite{HalatiKollath2020, HalatiKollath2022, HalatiKollath2025}.
We emphasize that our results cannot be inferred from considerations based on general bounds on the propagation of correlations, which would be dominated by the contributions of the global-range couplings, such that numerical simulations of the quantum dynamics are needed. 
The studied effects could be realized and investigated in current state-of-the-art experiments of ultra-cold atoms coupled to optical cavities. However, due to its generality our findings are applicable to a much wider class of experimental systems. Our results can guide the experimental investigations in controlling the spreading of correlations in platforms with competing interaction scales.

\emph{Acknowledgments:}
We thank J.P.~Brantut and T.~Donner for fruitful discussions.
We acknowledge support by the Swiss National Science Foundation under Division II Grant No.~200020-219400 and by the Deutsche Forschungsgemeinschaft (DFG, German Research Foundation) under Project No.~277625399-TRR 185 OSCAR (``Open System Control of Atomic and Photonic Matter'', B4), No.~277146847-CRC 1238 (``Control and dynamics of quantum materials'', C05), CRC 1639 NuMeriQS (``Numerical Methods for Dynamics and Structure Formation in Quantum Systems'') – project
No.~511713970 , Project-ID 429529648 – TRR 306 QuCoLiMa (“Quantum Cooperativity of Light and Matter’’), 
and under Germany’s Excellence Strategy – Cluster of Excellence Matter and Light for Quantum Computing (ML4Q) EXC 2004/1 – 390534769. 
We also acknowledge support within the QuantERA II Programme (project "QNet: Quantum transport, metastability, and neuromorphic applications in Quantum Networks"), which has received funding from the EU's Horizon 2020 research and innovation programme under Grant Agreement No.~101017733, as well as from the DFG (Project ID 532771420).
This research was supported in part by the National Science Foundation under Grants No.~NSF PHY-1748958 and PHY-2309135.

\emph{Data availability:} 
The supporting data for this article are openly available at Zenodo \cite{datazenodo}.

\pagebreak

\section{Supplemental Material}

\setcounter{section}{0}
\renewcommand{\thesection}{\Alph{section}}
\setcounter{section}{0}
\renewcommand{\thesubsection}{\arabic{subsection}}

\setcounter{equation}{0}
\renewcommand{\theequation}{A.\arabic{equation}}
\setcounter{figure}{0}
\renewcommand{\thefigure}{A\arabic{figure}}

\section{\label{app:method_numerics} Time-dependent matrix product states method for long-range quantum systems}

The numerically exact simulations for the time evolution of the Liouvillian of  a one-dimensional Bose-Hubbard model coupled to a dissipative cavity [Eqs.~(1)-(2) in the main text] have been performed by employing a recent implementation of a matrix product states (MPS) method \cite{HalatiKollath2020, HalatiKollath2020b, HalatiPhd}. The time-evolution of the Hamiltonian which contains the global-range coupling to the cavity makes use of a variant of the quasi-exact time-dependent variational matrix product state (tMPS) based on the Trotter-Suzuki decomposition of the time evolution propagator \cite{WhiteFeiguin2004, DaleyVidal2004, Schollwoeck2011} and the dynamical deformation of the MPS structure using swap gates \cite{StoudenmireWhite2010, Schollwoeck2011, WallRey2016}.
For the results in the presence of cavity losses we employ the stochastic unravelling of the master equation with quantum trajectories \cite{DalibardMolmer1992, GardinerZoller1992, Daley2014}.
The method has been implemented by employing the ITensor Library \cite{FishmanStoudenmire2020}. 
Additional details regarding the implementation and benchmarks can be found in Ref.~\cite{HalatiKollath2020b}.

In order to ensure the convergence of the density-density correlations, $\mathcal{C}_{nn}$, [see main text Eq.~(3) for their definition], up to an error of approximately $~10^{-5}$ for the times considered in the results presented in the absence of dissipation, we chose the following convergence parameters: a maximal bond dimension of $400$ states, which ensured a truncation error of at most $5\times10^{-8}$ at the final time, a time-step of $dt J/\hbar=2.5\times10^{-3}$, the local Hilbert space of the bosonic atoms is $N_\text{bos}=4$, and the adaptive cutoff of the local Hilbert space of the photonic mode ranged between $N_\text{pho}=70$ and $N_\text{pho}=8$. 
For the parameters used in the presence of dissipation a truncation error of $10^{-8}$ at the final time was achieved for the mentioned convergence parameters, the presented results are averaged over at least 750 quantum trajectories.

The numerical results for the time-evolution of the atomic model with global range interactions, Eq.~(4) in the main text, were obtained with an implementation based on matrix product states of the two-site version of the time-dependent variational principle approach (TDVP) \cite{HaegemanVerstraete2011, HaegemanVerstraete2016}, within the ITensor Library \cite{FishmanStoudenmire2020}. 
The convergence was assured by the following parameters of the method: a maximal bond dimension of $250$ states, which ensured a truncation error of at most $10^{-7}$ at the final time, a time-step of $dt J/\hbar=5\times10^{-3}$, and the local Hilbert space of the bosonic atoms are $N_\text{bos}=4$.

\setcounter{equation}{0}
\renewcommand{\theequation}{B.\arabic{equation}}
\setcounter{figure}{0}
\renewcommand{\thefigure}{B\arabic{figure}}

\section{Approximate Analytical calculations}

\subsection{\label{app:smallcoupling} In the limit of small coupling}

In this section, we approximate the time-evolution operator in the limit of a small coupling $\hbar\Omega \ll J, U$ and short times $t$.
The time-evolution operator is given by $\hat{U}(t)=e^{-i\hat{H}t/\hbar}$. Thus, for the Hamiltonian
\begin{align} 
\label{eq:Hamiltonian}
&\hat{H}=\hat{H}_{\text{c}}+\hat{H}_{\text{ac}}+\hat{H}_{\text{int}}+\hat{H}_{\text{kin}} \\
&\hat{H}_{\text{int}}=\frac{U}{2} \sum_{j} \hat{n}_{j}(\hat{n}_{j}-1),~\hat{H}_{\text{kin}}=-J \sum_{\langle j,j'\rangle} (\hat{b}_{j}^\dagger \hat{b}_{j'} + \text{H.c.}), \nonumber\\
&\hat{H}_{\text{c}}= \hbar\delta \hat{a}^\dagger \hat{a},~\hat{H}_{\text{ac}}=  -\hbar\Omega ( \hat{a} + \hat{a}^\dagger)\hat{\Delta},\nonumber
\end{align}
by employing the Zassenhaus formula we can rewrite the exponential of the sum of different terms as
\begin{align} 
\label{eq:expansion}
\hat{U}(t)^\dagger&=e^{i\hat{H}t/\hbar} \\
&= e^{i(\hat{H}_{\text{kin}}+\hat{H}_{\text{int}})t/\hbar}e^{i(\hat{H}_{\text{ac}}+\hat{H}_{\text{c}})t/\hbar}\nonumber\\
&\qquad\times e^{-t^2/2\hbar[\hat{H}_{\text{kin}}+\hat{H}_{\text{int}},\hat{H}_{\text{ac}}]} \dots, \nonumber
\end{align}
where the dots stand for exponentials of higher order commutators. 
By expanding the exponentials in the orders of $\Omega$ we obtain
\begin{align} 
\label{eq:expansion2}
\hat{U}(t) \approx  e^{-i(\hat{H}_{\text{kin}}+\hat{H}_{\text{int}})t/\hbar} +O(t\Omega).
\end{align}
The term linear in $t\Omega$ vanishes in the evolution of the density-density correlations, since the coupling term $\hat{H}_{ac}$ commutes with the density-density operator, i.e.
\begin{align} 
  \frac{\partial}{\partial t}&\langle \hat{n}_j \hat{n}_{j+d}\rangle\propto  \\
  &~\big\langle e^{i(\hat{H}_{\text{kin}}+\hat{H}_{\text{int}})t/\hbar} \left[1-i t\Omega  ( \hat{a} + \hat{a}^\dagger)\hat{\Delta}\right] \hat{n}_j \hat{n}_{j+d} \nonumber\\
  &~~\times\left[1+i t\Omega  ( \hat{a} + \hat{a}^\dagger)\hat{\Delta}\right] e^{-i(\hat{H}_{\text{kin}}+\hat{H}_{\text{int}})t/\hbar} \big\rangle+O(t^2\Omega)\nonumber\\
  =&
  \big\langle e^{i(\hat{H}_{\text{kin}}+\hat{H}_{\text{int}})t/\hbar}  \hat{n}_j \hat{n}_{j+d}  e^{-i(\hat{H}_{\text{kin}}+\hat{H}_{\text{int}})t/\hbar} \big\rangle +O(t^2\Omega) \nonumber
\end{align}
A similar argument holds for the evolution of the average value of the local density. 
If one takes these expression together, it implies that the time-evolution of the density-density correlations are dominated by the short range terms of the Hamiltonian and obtain only corrections on the order of $O(\Omega t^2)$ from the global-range couplings. 
For many different initial states, one can further show that only higher order correction are important in the short-time dynamics, as we show in the following section.

\subsection{Minimal model for the description of the short-time density-density correlation growth}

In this section, we aim to capture the minimal ingredients necessary for understanding the distance-independent spreading of the correlations and derive the short-time scaling for the minimal model, $\mathcal{C}_{nn}(d,t)\propto J^4 \Omega^4 t^8$. 
As discussed in the main text, when starting from an uncorrelated Fock state the kinetic processes are responsible for creating density-density correlations locally and the coupling to the cavity determines the supersonic spreading. Thus, as a first approximation we neglect the on-site repulsive interactions and the detuning $\delta$ to explain the short time behaviour.
In this situation, the 1D version of the Hamiltonian in Eq.~(2) of the main text can be written in the momentum basis as 
\begin{align}
	\hat{H}=-2J\sum_{k\in\mathbb{B}}\cos(k)\hat{b}^\dag_k\hat{b}_k- \hbar\Omega(\hat{a}+\hat{a}^\dag)\hat{\Delta},
\end{align}
where we assumed periodic boundary conditions and with 
\begin{align}
	\hat{\Delta}=&\sum_{k\in\mathbb{B}}\hat{b}_{k}^\dag\hat{b}_{k+\pi},\\
	\hat{b}_k=&\frac{1}{\sqrt{L}}\sum_{l}e^{ikl}\hat{b}_l,
\end{align}
and $\mathbb{B}=\{-\pi,-\pi+2\pi/L,\dots,\pi-2\pi/L\}$, where $L$ is the number of lattice sites.

This Hamiltonian can be rewritten in the form
\begin{align}
\label{eq:Ham_k}
	\hat{H}=\sum_{k\in\mathbb{B}'}\begin{pmatrix}
		\hat{b}^\dag_{k+\pi}&	\hat{b}^\dag_{k}
	\end{pmatrix}\begin{pmatrix}
		2J\cos(k)&-\hbar\hat{\Omega}\\
		-\hbar\hat{\Omega}&-2J\cos(k)
	\end{pmatrix}\begin{pmatrix}
	\hat{b}_{k+\pi}\\	\hat{b}_{k}
	\end{pmatrix}
\end{align}
with $\mathbb{B}'=\{-\pi/2+2\pi/L,-\pi/2+4\pi/L,\cdots,\pi/2\}$ and
\begin{align}
	\hat{\Omega}=&\Omega(\hat{a}^\dag+\hat{a}).
\end{align}
In the following we make an additional assumption by neglecting the full momentum dependence of the kinetic term and replacing it with the constant $J$ in the diagonal entries of Eq.~(\ref{eq:Ham_k}). This is justified in the regime in which the energy scale $\hbar\Omega$ is much larger than $J$.
The resulting Hamiltonian reads
\begin{align}
	\hat{H}\approx\sum_{k\in\mathbb{B}'}\begin{pmatrix}
		\hat{b}^\dag_{k+\pi}&	\hat{b}^\dag_{k}
	\end{pmatrix}\begin{pmatrix}
		J&-\hbar \hat{\Omega}\\
		-\hbar\hat{\Omega}&-J
	\end{pmatrix}\begin{pmatrix}
		\hat{b}_{k+\pi}\\	\hat{b}_{k}
	\end{pmatrix}.
	\end{align}
This simplified Hamiltonian can be diagonalized 
\begin{align}
	H=\sum_k\hat{E}[\hat{e}_{k,+}^\dag\hat{e}_{k,+}-\hat{e}_{k,-}^\dag\hat{e}_{k,-}],
\end{align}
with $\hat{E}=\sqrt{J^2+\hbar^2\hat{\Omega}^2}$ and
\begin{align}
	\hat{e}_{k,+}=\cos(\hat{\theta})\hat{b}_{k+\pi}-\sin(\hat{\theta})\hat{b}_{k},\label{eq:ek+}\\
	\hat{e}_{k,-}=\sin(\hat{\theta})\hat{b}_{k+\pi}+\cos(\hat{\theta})\hat{b}_{k},\label{eq:ek-}
\end{align}
where
\begin{align}
	\cos(\hat{\theta})=&\frac{J+\hat{E}}{\sqrt{2\hat{E}(\hat{E}+J)}},\\
	\sin(\hat{\theta})=&\frac{\hat{\Omega}}{\sqrt{2\hat{E}(\hat{E}+J)}}.
\end{align}

We use this result to derive the time-evolution of the bosonic annihilation and creation operators in real space.
In the case of $l$ is even we obtain
\begin{align}
		\hat{b}_l^\dag(t)=&\frac{1}{\sqrt{L}}\sum_{k\in\mathbb{B}'}e^{-ikl}[\hat{b}_{k+\pi}^\dag(t)+\hat{b}_{k}^\dag(t)]\nonumber\\
			=&\frac{1}{\sqrt{L}}\sum_{k\in\mathbb{B}'}e^{-ikl}\cos(\hat{\theta})[e^{i\hat{E}t/\hbar}\hat{e}_{k,+}^\dag+\hat{e}_{k,-}^\dag e^{-i\hat{E}t/\hbar}]\nonumber\\
		&-\frac{1}{\sqrt{L}}\sum_{k\in\mathbb{B}'}e^{-ikl}\sin(\hat{\theta})[e^{i\hat{E}t/\hbar}\hat{e}_{k,+}^\dag-\hat{e}_{k,-}^\dag e^{-i\hat{E}t/\hbar}]\nonumber\\
		=&\left[\cos(\hat{E}t/\hbar)-i\frac{\hbar\hat{\Omega}\sin(\hat{E}t/\hbar)}{\hat{E}}\right]\hat{b}_l^\dag(0),
		\end{align}
		where we used Eqs.~\eqref{eq:ek+} and \eqref{eq:ek-} and ${\sin(2\hat{\theta})=\hbar\hat{\Omega}/\hat{E}}$.
Consequently, we can calculate for two even sites $l,m$ the density-density correlations with
\begin{align}
  \label{eq:decoupling}
\mathcal{C}_{nn}(|l-m|,t)=&\left\langle\hat{n}_l\hat{n}_m\right\rangle-\left\langle\hat{n}_l\right\rangle\left\langle\hat{n}_m\right\rangle\nonumber\\
=&\mathrm{Var}(\hat{C})\langle\hat{n}_l(0)\rangle\langle\hat{n}_m(0)\rangle,
		\end{align}
with
\begin{align}
	\hat{C}=& \cos^2(\hat{E}t/\hbar)+\frac{\hbar^2\hat{\Omega}^2\sin^2(\hat{E}t/\hbar)}{\hat{E}^2},\\
	\mathrm{Var}(\hat{C})=&\langle\hat{C}^2\rangle-\langle\hat{C}\rangle^2.
\end{align}
Here we used that the initial state is a product of the photonic state and a Fock state of the atoms in order to decouple the expectation values in Eq.~(\ref{eq:decoupling}).

The final result now follows from the Taylor expansion of $\hat{C}$ for small times
\begin{align}
	\hat{C}\approx1-J^2t^2/\hbar^2+\frac{J^4t^4}{3\hbar^4}+\frac{J^2\hat{\Omega}^2t^4}{3\hbar^2}
\end{align}
and therefore
\begin{align}
\mathrm{Var}(\hat{C})&\approx\frac{J^4\Omega^4t^8}{9\hbar^2}\mathrm{Var}([\hat{a}+\hat{a}^\dag]^2),
\end{align}
This equation predicts the growth of density-density correlations  ${\mathcal{C}_{nn}\propto J^4\Omega^4 t^8}$ and shows the importance of the fluctuations of the cavity field operators.

\setcounter{equation}{0}
\renewcommand{\theequation}{C.\arabic{equation}}
\setcounter{figure}{0}
\renewcommand{\thefigure}{C\arabic{figure}}

\section{\label{app:classical_field} Treating the cavity mode as a classical field}

\begin{figure}[!hbtp]
\centering
\includegraphics[width=0.48\textwidth]{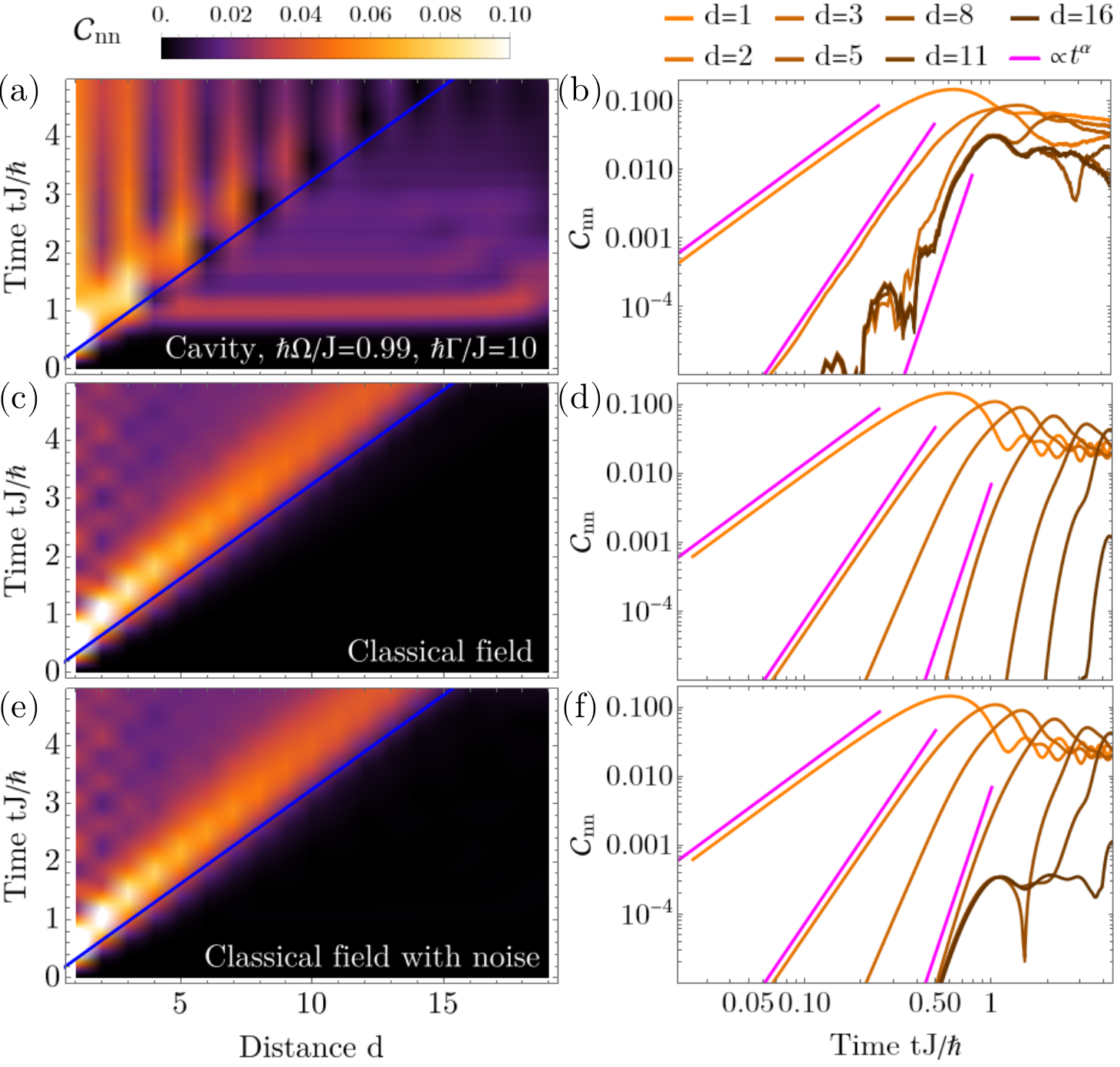}
\caption{(a), (c), (e) The space-time propagation of the correlations $\mathcal{C}_{nn}(d,t)$, in the presence of dissipation, and (b), (d), (f) the time-dependence of $\mathcal{C}_{nn}(d,t)$ for several distances.
In panels (a), (b) we show the full quantum dynamics of the atoms-cavity model, Eqs.~(1)-(2) in the main text, in panels (c), (d) the cavity dynamics has been replaced with a classical field, Eq.~(\ref{eq:Hamiltonian_MF}) and Eq.~(\ref{eq:mf_photon_field}), and in panels (e), (f) stochastic noise has been added to the classical field evolution, Eq.~(\ref{eq:Hamiltonian_MF}) and Eq.~(\ref{eq:mf_photon_field_noise}).
The blue lines are a guide to the eye and approximates the front of the light-cone propagation for $\Omega=0$ and $\Gamma=0$.
The magenta lines represent algebraically increasing curves $\propto t^\alpha$, with $\alpha\in\{2,4,8\}$.
The parameters used are $\hbar\Omega/J=0.99$ $\hbar\Gamma/J=10$, $N=10$ particles, $L=20$ sites, $U/J=2$, $\hbar\delta/J=2$.
Panels (a) and (c) correspond to Fig.~3(a) and Fig.~3(b) in the main text, we reproduce them here for completeness.
}
\label{fig:dissipation_supp}
\end{figure}

\begin{figure}[!hbtp]
\centering
\includegraphics[width=0.48\textwidth]{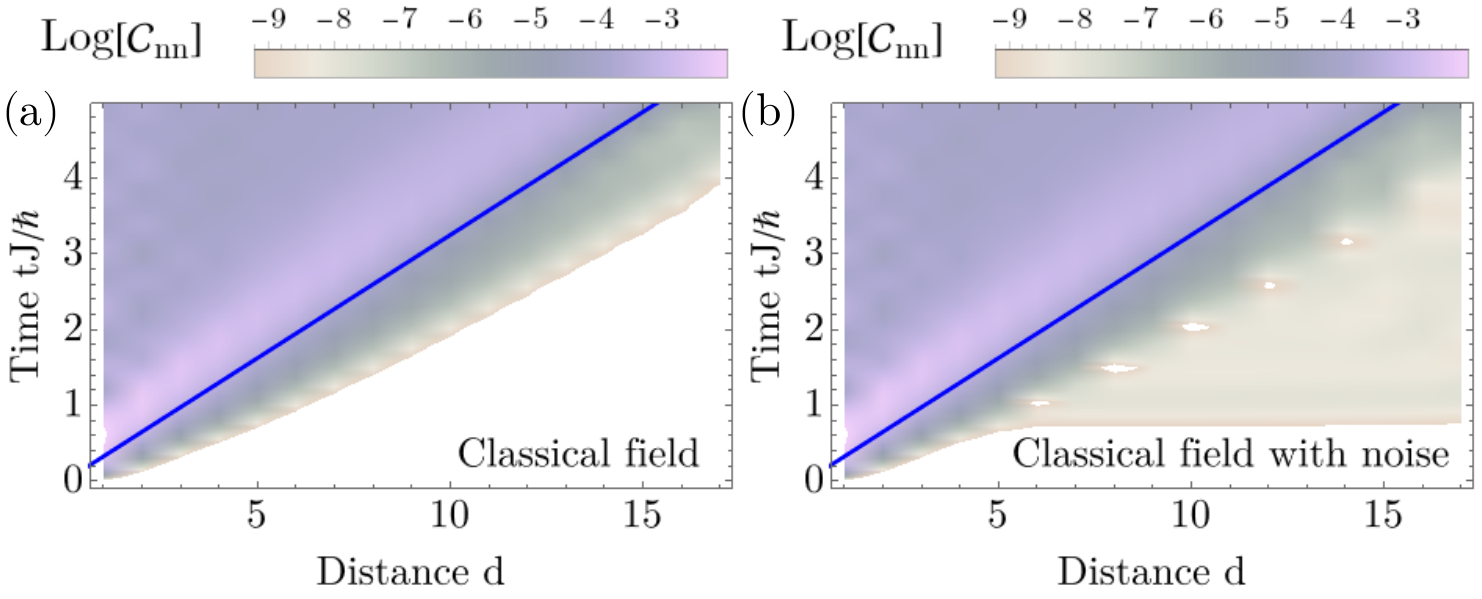}
\caption{
The space-time propagation of the correlations $\text{Log}[\mathcal{C}_{nn}(d,t)]$, in the presence of dissipation, where in (a) the cavity dynamics has been replaced with a classical field, Eq.~(\ref{eq:Hamiltonian_MF}) and Eq.~(\ref{eq:mf_photon_field}), and in panel (b) stochastic noise has been added to the classical field evolution, Eq.~(\ref{eq:Hamiltonian_MF}) and Eq.~(\ref{eq:mf_photon_field_noise}).
The panels show the same data as in Fig.~\ref{fig:dissipation_supp}(c) and Fig.~\ref{fig:dissipation_supp}(e), but in a logarithmic scale.
The blue lines are a guide to the eye and approximates the front of the light-cone propagation for $\Omega=0$ and $\Gamma=0$.
In the white regions the value of the correlations is smaller than $10^{-4}$.
The parameters used are $\hbar\Omega/J=0.99$ $\hbar\Gamma/J=10$, $N=10$ particles, $L=20$ sites, $U/J=2$, $\hbar\delta/J=2$.
}
\label{fig:dissipation_log}
\end{figure}

In the main text we contrast the propagation dynamics of the correlations for the Bose-Hubbard model coupled to the quantum dissipative field of an optical cavity with results in which the cavity mode has been replaced with a classical field. The classical field realizes a superlattice staggered potential.

The classical staggered potential can be derived as a mean-field description of the cavity-atoms coupling approach in which the cavity field is described by a classical coherent field \cite{RitschEsslinger2013, MaschlerRitsch2008, NagyDomokos2008}.
For the one-dimensional case, within this approximation, the atoms are described by the Hamiltonian
\begin{align} 
\label{eq:Hamiltonian_MF}
&\hat{H}_\text{MF}=\hat{H}_{\text{int}}+\hat{H}_{\text{kin}}+\hat{H}_{\text{stag}} \\
&\hat{H}_{\text{int}}=\frac{U}{2} \sum_{j=1}^L \hat{n}_{j}(\hat{n}_{j}-1),\nonumber\\
&\hat{H}_{\text{kin}}=-J \sum_{j=1}^{L-1} (\hat{b}_{j}^\dagger \hat{b}_{j+1} + \hat{b}_{j+1}^\dagger \hat{b}_{j}), \nonumber\\
&\hat{H}_{\text{stag}}=  -V(t) \hat{\Delta}, ~~ \hat{\Delta}=\sum_{j=1}^L (-1)^j \hat{n}_j \nonumber.
\end{align}
Here $V(t)$ is a time-dependent field, describing the coupling to cavity mode which is assumed to be in a coherent state,
$V(t)=\Omega \left\langle \hat{a}^\dagger+ \hat{a}\right\rangle(t)$, with the equation of motion
\begin{align} 
\label{eq:mf_photon_field}
&\frac{\partial}{\partial t} \langle \hat{a} \rangle = i \Omega \langle\hat{\Delta}\rangle-\left(i\delta+\Gamma/2\right)\langle\hat{ a} \rangle.
\end{align}
We integrate this equation of motion in a coupled way with the time-evolution of $H_\text{MF}$, where the potential coupled to the atomic odd-even imbalance, $V(t)$, depends on the imbalance $\langle\hat{\Delta}\rangle (t-\text{d}t)$ at the previous time step.

The comparison between the exact results and the classical field approach is also shown in Fig.~\ref{fig:dissipation_supp}.
As discussed in the main text, by the coupling to the cavity field for the parameters shown the dynamics of the correlations is dominated by a supersonic propagation [Fig.~\ref{fig:dissipation}(a) and Fig.~\ref{fig::dissipation_supp}(b)], while for the classical field approach we only observe a light-cone spreading of the correlations [Fig.~\ref{fig:dissipation_supp}(c) and Fig.~\ref{fig:dissipation_supp}(d)].

In order to understand the necessary ingredients for the supersonic propagation of the correlations, we also contract our results with the classical field approach for describing the cavity field to which stochastic noise has been added to its dynamics
\begin{align} 
\label{eq:mf_photon_field_noise}
&\frac{\partial}{\partial t} \langle \hat{a} \rangle = i \Omega \langle\hat{\Delta}\rangle-\left(i\delta+\Gamma/2\right)\langle\hat{ a} \rangle+\sqrt{\Gamma}\xi(t),
\end{align}
where $\xi(t)$ is a random complex number of magnitude $1$ sampled from a uniform distribution at each point in time. For the initial state $\langle a \rangle (0)$ we use a uniformly sampled random complex number. We simulate different realizations of the time-dependent noise term, similar to the quantum trajectories approach, and average over the values of the computed observables for the different realizations.
We obtain that in the presence of the stochastic noise, Fig.~\ref{fig:dissipation_supp}(e) and Fig.~\ref{fig:dissipation_supp}(f) it exhibits for larger distances features of the distance-independent rise of the correlations, as seen in Fig.~\ref{fig:dissipation_supp}(f) and in the log-scale depiction of the correlations in Fig.~\ref{fig:dissipation_log}. 
This can be understood by the fact that the noise term appearing in Eq.~(\ref{eq:mf_photon_field_noise}) translates to a correlated noise term in the evolution of the atoms acting on all sites. However, the contribution from the distance-independent rise of the correlations is much smaller for the shown parameters compared to the full quantum dynamics. This implies that the noise term is insufficient to capture the full dynamics which requires, in particular, also the cavity-mediated transport of density fluctuations.

\setcounter{equation}{0}
\renewcommand{\theequation}{D.\arabic{equation}}
\setcounter{figure}{0}
\renewcommand{\thefigure}{D\arabic{figure}}

\begin{figure}[!hbtp]
\centering
\includegraphics[width=0.48\textwidth]{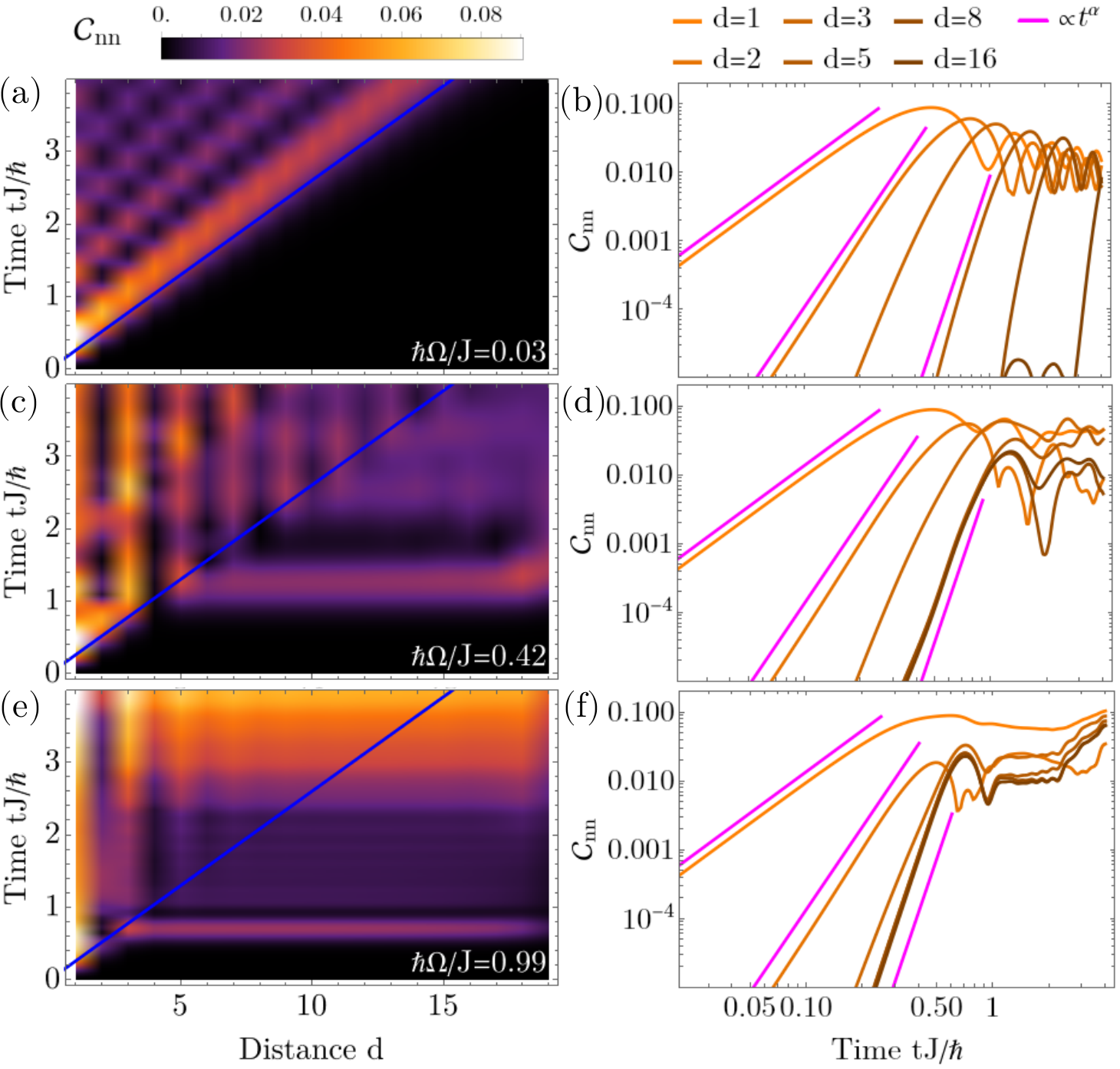}
\caption{
(a), (c), (e) The propagation of $\mathcal{C}_{nn}(d,t)$, computed with $\hat{H}$, Eq.~\eqref{eq:Hamiltonian}, for the atoms-cavity coupling strength $\hbar\Omega/J\in\{0.03,0.42,0.99\}$ and $U/J=16$. The blue line is a guide to the eye and approximates the front of the light-cone propagation for $\Omega=0$.
(b), (d), (f) The dynamics of $\mathcal{C}_{nn}(d,t)$ for several distances and $\hbar\Omega/J\in\{0.03,0.42,0.99\}$.
The magenta lines represent algebraically scaling $\propto t^\alpha$, with $\alpha\in\{2,4,8\}$.
The parameters used are $N=10$ particles, $L=20$ sites, $\hbar\delta/J=2$, $\hbar\Gamma/J=0$.
}
\label{fig:lightcone_U16}
\end{figure}

\section{\label{app:interaction} Role of on-site interactions}

In this section, we show the propagation of correlations in the regime of strong on-site interactions. In Fig.~\ref{fig:lightcone_U16} we show the equivalent of Fig.~1 of the main text for $U/J=16$ instead of $U/J=2$.
We see the same qualitative behavior as described in the main text, with a light-cone propagation of the correlations for small atoms-cavity coupling [see Fig.~\ref{fig:lightcone_U16}(a)] followed by a crossover to supersonic, distance-independent, propagation at larger values of the coupling [see Fig.~\ref{fig:lightcone_U16}(c),(e)].
The value of the atomic on-site interactions does not play a crucial role in the observed phenomenology as they commute with the global-range terms, thus, for the chosen initial state the kinetic terms are the ones which
seed the initial correlations, transported afterwards by the global couplings.
However, the on-site interactions are important for the details of the dynamics, e.g.~the space and time oscillations observed in Fig.~\ref{fig:lightcone_U16} and the velocity of the light-cone propagation \cite{CheneauKuhr2012}.

\setcounter{equation}{0}
\renewcommand{\theequation}{E.\arabic{equation}}
\setcounter{figure}{0}
\renewcommand{\thefigure}{E\arabic{figure}}

\section{\label{app:method_TWA} Truncated Wigner approximation (TWA)}

In this section, we describe the truncated Wigner simulation methods \cite{GardinerFudge2002,OlsenCavalcanti2004,NagaoDanshita2019} and we provide additional information regarding the truncated Wigner results presented in the main texts. 
The truncated Wigner simulation is based on semiclassical equations of motion for the bosonic field operators describing the atomic and the cavity degrees of freedom. 
For each lattice site $j$ we have two real fields $\beta^{(r)}_j$ and $\beta^{(i)}_j$ that can be captured as one complex field $\beta_j=\beta^{(r)}_j+i\beta^{(i)}_j$ representing the bosonic operator $\hat{b}_j$. In addition, we have two real cavity fields $\alpha^{(r)}$ and $\alpha^{(i)}$ that define a complex cavity field $\alpha=\alpha^{(r)}+i\alpha^{(i)}$ representing the bosonic cavity field operator $\hat{a}$. 

In order to derive the semiclassical equations of motions we start with the Heisenberg-Langevin equations of motion for the operators that are given by
\begin{align}
	\frac{d\hat{b}_j}{dt}=&i\frac{J}{\hbar}\sum_{l\in \langle l,j\rangle}\hat{b}_l-i\frac{U}{\hbar}\hat{b}_j^\dag\hat{b}_j\hat{b}_j\nonumber\\
	&+i\Omega(\hat{a}+\hat{a}^\dag)(-1)^j\hat{b}_j\\
	\frac{d\hat{a}}{dt}=&-\left(i\delta+\frac{\Gamma}{2}\right)\hat{a}+\sqrt{\Gamma}\hat{a}_{\mathrm{in}}(t)\\
	&+i\Omega\sum_{j}(-1)^j\hat{n}_j\nonumber,
\end{align}
where $\langle l,j\rangle$ is the set of neighboring sites of $j$.
The input shot noise $\hat{a}_{\mathrm{in}}(t)$ has vanishing mean value $\langle \hat{a}_{\mathrm{in}}(t)\rangle$ and second moments $\langle\hat{a}_{\mathrm{in}}(t)\hat{a}_{\mathrm{in}}(t')\rangle=0=\langle\hat{a}^\dag_{\mathrm{in}}(t)\hat{a}_{\mathrm{in}}(t')\rangle$, $\langle\hat{a}_{\mathrm{in}}(t)\hat{a}^\dag_{\mathrm{in}}(t')\rangle=\delta(t-t')$. 

To obtain the semiclassical equations of motion for the real fields $\alpha^{(r)}$, $\alpha^{(i)}$ and $\beta^{(r)}_j$, $\beta^{(i)}_j$, we first derive the equations of motion for $\hat{b}_j^{(r)}=(\hat{b}_j+\hat{b}_j^\dag)/2$, $\hat{b}_j^{(i)}=(\hat{b}_j-\hat{b}_j^\dag)/(2i)$ and $\hat{a}^{(r)}=(\hat{a}+\hat{a}^\dag)/2$, $\hat{a}^{(i)}=(\hat{a}-\hat{a}^\dag)/(2i)$. Subsequently we perform a symmetric ordering of the operators. Note that for instance
$\hat{n}_j=\hat{b}_j^\dag\hat{b}_j=[\hat{b}_j^{(r)}]^2+[\hat{b}_j^{(i)}]^2-1/2$. After that we exchange the operators $\hat{b}_j^{(r)}$, $\hat{b}_j^{(i)}$ by the real variables $\beta^{(r)}_j$, $\beta^{(i)}_j$ and $\hat{a}^{(r)}$, $\hat{a}^{(i)}$ by the real variables $\alpha^{(r)}$, $\alpha^{(i)}$. With these equations of motion we can write down a set of complex coupled stochastic differential equations for $\alpha=\alpha^{(r)}+i\alpha^{(i)}$ and $\beta_j=\beta^{(r)}_j+i\beta^{(i)}_j$ given by
\begin{align}
	\frac{d\beta_j}{dt}=&i\frac{J}{\hbar}\sum_{l\in n(j)}\beta_l-i\frac{U}{\hbar}\left(|\beta_j|^2-\frac{1}{2}\right)\beta_j\label{eq:beta}\\
	&+i\Omega(\alpha+\alpha^*)(-1)^j\beta_j\nonumber\\
	\frac{d\alpha}{dt}=&-\left(i\delta+\frac{\Gamma}{2}\right)\alpha+\sqrt{\Gamma}\mathcal{F}(t)\label{eq:alpha}\\
	&+i\Omega\sum_{j}(-1)^j\left(|\beta_j|^2-\frac{1}{2}\right).\nonumber
\end{align}
with $\mathcal{F}(t)=[\mathcal{F}_r(t)+i\mathcal{F}_i(t)]/2$, $\langle\mathcal{F}_r\rangle=0=\langle\mathcal{F}_i\rangle$, $\langle\mathcal{F}_r(t')\mathcal{F}_i(t)\rangle=0$, and $\langle\mathcal{F}_i(t')\mathcal{F}_i(t)\rangle=\delta(t-t')$, $\langle\mathcal{F}_r(t')\mathcal{F}_r(t)\rangle=\delta(t-t')$.
Equations~\eqref{eq:beta} and \eqref{eq:alpha} are the dynamical equations we simulate in our truncated Wigner approach with a given initial condition. 

\subsection{Initial conditions}

 To model the initial condition of the atomic Fock state and the empty cavity states used in the simulations of the main text, we initialize the cavity in the vacuum state and sample $\alpha^{(r)}(t=0)$ and $\alpha^{(i)}(t=0)$ with $\alpha(t=0)=\alpha^{(r)}(0)+i\alpha^{(i)}(0)$ as Gaussian random variables with $\langle\alpha^{(r)}(0)\rangle=0=\langle\alpha^{(i)}(0)\rangle$, $\langle\alpha^{(r)}\alpha^{(i)}\rangle=0$, and $\langle\alpha^{(r)}(0)\alpha^{(r)}(0)\rangle=1/4=\langle\alpha^{(i)}(0)\alpha^{(i)}(0)\rangle$. For the atomic state we assume that each mode indexed by $j$ is in a Fock state with occupation $n_j$. For this we employ a sampling method which was previously derived in Refs.~\cite{GardinerFudge2002,OlsenCavalcanti2004,NagaoDanshita2019}. We sample $\beta_j(0)=[p_j+q_j\eta_j]e^{i\varphi_j}$ where $\eta_j$ is a Gaussian random variable with mean equal zero and with variance $\langle\eta_j^2\rangle=1$ and $\varphi_j$ is a uniform random variable in the interval $[0,2\pi)$. The values of $p_j$ and $q_j$ are determined by the occupation $n_j$ with
\begin{align}
	p_j=&\frac{\sqrt{2n_j+1+2\sqrt{n_j^2+n_j}}}{2},\\
	q_j=&\frac{1}{4p_j}.
\end{align} 
Note that with this definition we find
\begin{align}
	\langle\beta_j^*(0)\beta_j(0)\rangle=&n_j+\frac{1}{2}
\end{align}
and
\begin{align}
	\langle\beta_j^*(0)\beta_j^*(0)\beta_j(0)\beta_j(0)\rangle
	=&\left(n_j+\frac{1}{2}\right)^2+\frac{1}{4}.
\end{align}

\subsection{Technical details on the simulations}

For the simulation of the stochastic differential equation we have used a Runge-Kutta method of 4th order with an integration time step $dt=0.5\times10^{-2}/\Gamma$. 
All results shown in the main text are obtained from $M=10^6$ noise initializations. For each of the initialization that we index in the following by $m=1,...,M$ we obtain fluctuating densities
\begin{align}
	n^{(m)}_j(t)=|\beta^{(m)}_j(t)|^2-\frac{1}{2}.
\end{align}
The approach is then slighty different for the 1D and 2D cases.
\paragraph{1D simulation:}
For the 1D simulation we choose $j=1,\dots,L$ with $L=21$ and we calculate the density-density correlations with
\begin{align}
  \mathcal{C}_{nn}(d,t)=&\frac{1}{M}\sum_{m=1}^{M}n^{(m)}_{j_0}
  (t)\frac{n^{(m)}_{ j_0+d}(t)+n^{(m)}_{ j_0-d}(t)}{2}\\
  &-\frac{1}{M^2}\sum_{m,m'=1}^{M}n^{(m')}_{j_0}(t)
\frac{n^{(m)}_{j_0+d}(t)+n^{(m)}_{j_0-d}(t)}{2}
\end{align}
where $j_0=(L+1)/2$ 
is the center and the division by $2$ comes from the fact that there are two sites with the same distance in 1D. The result of this calculation is shown in Fig.~3(c) in the main text. 
\paragraph{2D simulation:}
For the 2D simulation we use ${\bf j}=(j_x,j_y)$ with $j_x=1,\dots,L$ and $j_y=1,\dots,L$ and $L=21$. The center is then $\mathbf{j}_0=([L+1]/2,[L+1]/2)$. 

We can then calculate the density-density correlations with
\begin{align}
	\mathcal{C}_{nn}(d,t)=&\frac{1}{M}\sum_{m=1}^{M}n^{(m)}_{\mathbf{j}_0}(t)\sum_{{\bf j}}\frac{n^{(m)}_{{\bf j}}(t)}{\mathcal{N}_d}\delta_{d({\bf j}),d}\\
	&-\frac{1}{M^2}\sum_{m'=1}^{M}n^{(m')}_{\mathbf{j}_0}(t)\sum_{m=1}^{M}\sum_{{\bf j}}\delta_{d({\bf j}),d}\frac{n^{(m)}_{{\bf j}}(t)}{\mathcal{N}_d}\nonumber,
\end{align}
where $d({\bf j})=\sqrt{(j_x-[L+1]/2)^2+(j_y-[L+1]/2)^2}$ is the distance from the center and $\mathcal{N}_{d}$ is the number of sites with distance $d({\bf j})=d$. The result of this calculation is shown in Fig.~3(d) of the main text.

\setcounter{equation}{0}
\renewcommand{\theequation}{F.\arabic{equation}}
\setcounter{figure}{0}
\renewcommand{\thefigure}{F\arabic{figure}}

\section{\label{app:dephasing} Local dissipation} 

\begin{figure}[!hbtp]
\centering
\includegraphics[width=0.48\textwidth]{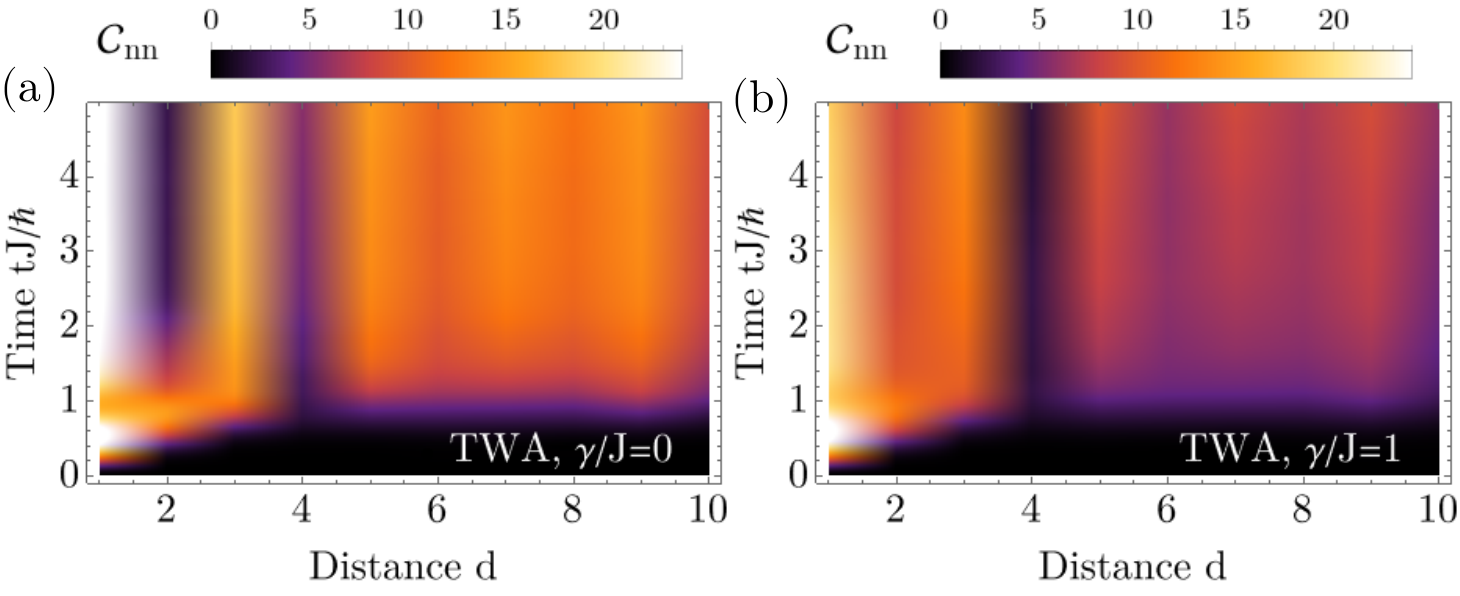}
\caption{Correlations $\mathcal{C}_{nn}(d,t)$ calculated from TWA simulations for (a) vanishing local dephasing $\gamma=0$ and (b) $\hbar\gamma/J=1$ in 1D for a lattice with $L=21$ sites. We simulated Eq.~\eqref{eq:beta} added Eq.~\eqref{eq:local} and Eq.~\eqref{eq:alpha} with $U/J=0.1$, $\hbar\delta/J=2$, $\hbar\Gamma/J=1$, $\sqrt{N}\hbar\Omega/J=4$. The atoms are initialized with alternating densities of $n=10$ and $n=11$ bosons corresponding to total atom number $N=220$ and the cavity is initialized in the vacuum state.
}
\label{fig:dephasing}
\end{figure}

In this section we show briefly that supersonic spreading can also be observed in presence of local atomic dissipation. To study the effect of local dissipation we use the TWA simulation introduced in the previous section.

Local dissipation is added to our previous model in form of local dephasing
\begin{align}
\mathcal{L}_j\hat{\rho}=&-\frac{\gamma}{2}(\hat{n}_j^2\hat{\rho}+\hat{\rho}\hat{n}_j^2-2\hat{n}_j\hat{\rho}\hat{n}_j)
\end{align}
 with rate $\gamma$ and such that the master equation governing the dynamics of the density operator $\hat{\rho}$ reads
\begin{align}
	\frac{\partial\hat{\rho}}{\partial t}=-&\frac{i}{\hbar}\left[\hat{H},\hat{\rho}\right]+\frac{\Gamma}{2}(2\hat{a}\hat{\rho}\hat{a}^\dag-\hat{a}^\dag\hat{a}\hat{\rho}-\hat{\rho}\hat{a}^\dag\hat{a})\nonumber\\
	+&\sum_j\mathcal{L}_j\hat{\rho}.
\end{align}
Employing symmetric ordering we can derive the dynamics for the complex-valued classical fields $\beta_j=\beta_j^{(r)}+i\beta_j^{(i)}$. The local dephasing term results in a term which needs to be added to Eq.~\eqref{eq:beta} which is
\begin{align}
	\left(\frac{d\beta_j}{dt}\right)_{\mathrm{local}}=-\frac{\gamma}{2}\beta_j -i\sqrt{\gamma}\beta_j W_j.\label{eq:local}
\end{align}
Here we introduced local noise terms $W_j$ that obey $\langle W_j(t)\rangle=0$ and $\langle W_j(t)W_i(t')\rangle=\delta_{ij}\delta(t-t').$

We simulate the corresponding coupled stochastic differential equation of the complex-valued fields $\alpha$ and $\beta_j$ for the same parameters as in Fig.~3(c) in the main text. In Fig.~\ref{fig:dephasing}(a) and (b) we show the results of the simulations for vanishing $\gamma=0$ and for non-vanishing $\hbar\gamma/J=1$, respectively. 
We find slight modifications in the magnitude of the correlations. In particular, the value of $\mathcal{C}_{nn}$ seems to be slightly smaller for $\hbar\gamma/J=1$ than it is for $\gamma=0$. In general, however, the shape of the correlation spreading is very similar for both cases. We highlight that we observe the light-cone like spreading for short times and the simultaneous supersonic spreading afterwards. This finding confirms that the features that are discussed in the paper are robust against (weak) local dephasing terms.

\end{document}